\documentstyle[12pt,psfig]{article}
\begin{document}
\sloppy
\newpage
%\pagestyle{myheadings}
%\markright{YPC-97}
\pagestyle{plain}
\def\be{\begin{equation}}
\def\ee{\end{equation}}

\noindent{\Large{\bf Electrical Conduction in Nonlinear Composites}}
\vskip 0.5cm
\begin{flushleft}
\parbox{9.0in}{\sf Abhijit Kar Gupta \& Asok K. Sen}\\
{\it LTP Section, Saha Institute of Nuclear Physics}\\
{\it 1/AF Bidhannagar, Calcutta 700 064, India}\\
\end{flushleft}
\vskip 0.2 in
\noindent {\large{\bf Introduction}}
\vskip 0.1 in

Composite systems constitute a large class of naturally occurring or 
artificially synthesized disordered systems \cite{exp}. 
The systems are microscopically inhomogeneous and disordered but
look homogeneous on the macroscopic scale. 
From the tunnelling electron micrographs
(TEM) of such a composite material it can be seen that the typical 
dimension ($\xi$) of 
metallic islands embedded in the insulating matrix are much greater 
than the atomic size ($a$) but obviously much smaller than the macroscopic 
scale length ($L$):~ $a \ll \xi \ll L$. 
The effective conductivity of such a system
depends upon the conductivities of the individual phases. 
For a low volume fraction ($p$) of the conducting phase, the system as a 
whole behaves like an insulator since the conducting regions do not form a 
continuous path through the sample. As $p$ is increased, the conducting 
regions will in general tend to grow and eventually at a critical volume 
fraction ($p_c$, called the percolation threshold) the conducting phase 
{\it percolates} through the sample. 
This may be considered as a classical insulator-to-metal transition or 
more popularly as a {\it percolation transition}. 
For all $p > p_c$, the system is metallic, and if the conducting phase 
is Ohmic, so is the whole macroscopic system. Clearly this class of systems 
may be well described by the geometrical percolation theory.

Now if an external voltage is applied across such composite systems
(examples include dispersed metallic systems, carbon-black-polymer 
composites, sulphonated (doped) polyaniline networks {\em etc.}, 
which are usually highly structured and give rise to some sort of universal 
behaviour.) a wide variety of {\bf interesting features} associated with
a nonlinear response emerge.  Usually these composites exhibit an unusually
{\it low percolation threshold}. Qualitatively {\it identical nonlinear
$I-V$ (as well as $dI/dV$ ($\equiv G$)against 
$V$)} response have been reported \cite{carbon, chen} both below and above 
the threshold for many of the composites although the nonlinearity exponent 
is found to be grossly different in the two regimes. Power-law  growth  of 
excess conductance for small $V$ is another general feature of the class of 
composite systems where non-integer power-law has 
been observed. This in turn implies a power-law in the $I-V$ relationship for 
small applied voltage ($V$). 
The $G-V$ curves are seen to {\it saturate} for an 
appropriately high enough voltage below the Joule-heating regime. The typical 
curve then looks like a nonlinear sigmoidal type function interpolating two 
linear regimes. 
Recent experiments on carbon-wax systems
\cite{carbon} as well as many earlier ones on disordered/ amorphous
systems \cite{hop}, find a non-integer power-law behaviour and a saturation in 
the DC-response as mentioned above.
Composite systems show very interesting temperature-dependent conduction 
properties particularly in the low temperature regime where the conduction 
is mainly due to phonon-assisted hopping (Mott's variable range hopping (VRH)). 
Some recent experiments indicate an effect of dilution on the relevant
temperature-exponent for fitting the low-temperature data with VRH or its other 
variations. We address many of the above mentioned features through our study.

\vskip 0.2 in
\noindent{\large{\bf Modelling Nonlinear Transport}}
\vskip 0.1 in

The framework of our study is based on percolation theory.
The ultra-low percolation threshold and the fact that many of these
nonlinear systems carry current even below $p_c$ indicates strongly
that {\it tunnelling} through disconnected (dispersed) metallic regions
must give some virtually connected percolating clusters.  
From the nonlinear $I-V$ characteristics ({\em e.g.}, see the experiment
by Chen and Johnson \cite{chen}) it is observed that the response (DC) behaviour 
is reversible with respect to the applied field. Also the temperature-dependent resistance with a 
minimum at some characteristic temperature and the Mott variable range hopping 
(VRH) type behaviour at very low temperatures give further credence to 
tunnelling assisted percolation. In practice, the tunnelling conductance 
should fall off exponentially with distance and hence the tunnelling should 
have an upper cut-off length scale. 
So for simplicity and to capture the basic physics, we construct a bond 
(lattice) percolation model for this problem, such that tunnelling may 
take place only between two nearest neighbour ohmic bonds
and no further. For a further simplification, we assume the nonlinear 
response of each tunnelling bond (or resistor) to be piecewise linear. 
We assume that all the tunnelling bonds 
have an identical voltage threshold ($v_g$) below which they are 
perfect insulators and above which they behave as ohmic conductors.
Clearly this is the source of nonlinearity in the model.
Made of both random resistive and tunnelling elements, this network will be 
called a random resistor cum tunnelling-bond network (RRTN) \footnote{In this 
respect we comment that a dynamic random resistor (DRRN) model proposed by 
Gefen {\em et al.}, \cite{exp} is different from our RRTN model in the sense 
that they allowed any insulating bond at any position in the lattice to break 
and turn metallic, whereas in our case such breakings can occur only at some 
correlated bond positions. Moreover, with the addition of these bridge bonds 
(anywhere), the new percolation threshold may not be properly defined like 
that of ours.}.

The {\bf percolation statistics} \cite{kgs} of the model network is examined in the 
saturation limit, {\em i.e.}, when all the tunnelling bonds can overcome their 
threshold. We estimate the new percolation ($p_{ct}$) threshold and address the 
question of universality class.
We undertake small-cell renormalization, Monte Carlo simulation and 
finite size scaling analysis to estimate $p_{ct}$ and some of the idependent 
critical exponents around it. The simulation results are obtained for lattices 
in 2D for convenience. Lattice sizes $L$ = 20 to 300 are considered.
The $p_{ct}$ is found  to be 0.181 $\pm$ 0.001 and the value of correlation
length exponent $\nu$ is obtained to be $\cong$ 1.35 $\pm$ 0.06. 
The {\it fractal dimension} ($D$) for the spanning 
cluster at the threshold which is found to be $D \cong$ 1.87, very 
close to 91/48, the fractal dimension for 2D random bond percolation. 
We also calculate the {\it conductivity exponent}, $t$, in the upper linear
regime where all the tunnelling resistors are considered to be behaving as 
the other ohmic resistors. 
We obtain $t/\nu \cong 0.90$, where the value of this ratio for the usual
percolation problem is $\cong 0.97$.
The value of the critical exponents, as obtained above, indicate  
that this correlated model for percolation belongs to the same 
{\it universality class} as that of its uncorrelated version (in the 
absence of tunnelling bonds).

An {\bf effective medium approximation (EMA)} \cite{kirk} has been used to 
calculate the percolation thereshold for our model system and the conductivity 
behaviour in the saturation limit. 
The probability of a bond to be ohmic, tunnelling or purely insulating 
according to the considerations of our model is:
$P_{ohm} = p$, $P_{tun} = (p^3+3p^2q+3pq^2)^2q$, $P_{ins} = 
[1-(p^3+3p^2q+3pq^2)^2]q$, where $q=1-p$.
If the conductances of the three types of bonds are denoted by 
$g_{ohm}$, $g_{tun}$ and $g_{ins}$, the EMA equation for this general 
situation can be written as
\be
{P_{ohm}(G_e - g_{ohm}) \over [g_{ohm} + (d-1)G_e]} + {P_{tun}(G_e - g_{tun}) 
\over [g_{tun} + (d-1)G_e]} + {P_{ins}(G_e - g_{ins}) \over [g_{ins} + 
(d-1)G_e]} = 0.   \label{emaeq} 
\ee
Solving the above equation for 2D ($d=2$) and 3D ($d=3$) we obtain 
$p_{ct}$ = 1/4 and 1/8 respectively. It may be noted that the value of 
$p_{ct}$ in 2D is close to that obtained by numerical simulation.
We examine the behaviour of the 
effective linear conductance ($G_e$) for the macroscopic model composite 
system in the saturation limit, given some specific values or forms of the 
resistive elements. These then are compared with the results obtained with 
the numerical simulation. The agreement is fairly good when one is away 
from the threshold, $p_{ct}$.

Next we study the {\bf dielectric breakdown} phenomenon in our model as 
the onset of nonlinear conduction against applied field for $p \le p_c$.
Below the percolation threshold ($p_c$) there exists a number of metallic 
clusters, isolated from each other, but closely spaced.
As new conducting paths are created when the local electric field 
across tunnelling bonds increases above $v_g$, the conductivity of the 
whole system jumps from a zero to a non-zero value (for $p < p_c$) as the
external applied field crosses \cite{skg}
the dielectric breakdown field ($E_B^p = V_B^p/L$). 
Note however that below $p_{ct}$ there is no sample-spanning cluster of 
combined ohmic and tunnelling  bonds, and hence there is no
breakdown at any finite electric field according to the criterion set for
our model. The interest would be to estimate the breakdown exponent $t_B$, 
where $E_B^p \sim (p_c - p)^{t_B}$.
To remove finite size effects, we work with the asymptotic breakdown field 
$E_B^p(L=\infty)$. From the least-square fit of the data for the above we find 
that the breakdown exponent $t_B \cong 1.42$ for our RRTN model.
It seems that the above exponent $t_B$ is not very different from that of the
usual breakdown exponent $t_B=\nu=$1.33 as discussed above. But it is not
unlikely either that we have a different result in our hands. If different, it 
could be because of the nature of the electric field in increasing the 
effective volume fraction of the conductors. 

We present the {\bf nonlinear DC-response} \cite{skg} namely, the 
current-voltage ($I-V$) and the conductance-voltage $G-V$ 
charactersistics in our model system. 
Our computer simulation involves solving Kirchhoff's law of current at the 
nodes of the RRTN network in 2D with the linear and nonlinear 
(assumed piecewise linear) resistors and the standard Gauss-Seidel relaxation 
technique.
We obtain current ($I$) and therefrom the differential conductance ($G = dI/dV$)
for the whole network at a given volume fraction $p$ of the ohmic bonds.
Simulation results for nonlinear $I-V$ curves for a square network of size 
$L=20$ were plotted in {\bf fig.~1} for $p$ = 0.3, 0.5 and 0.9. 
Averages over 50 configurations are done in each case.
One may note that the nonlinearity in the response exists for all $p$ both below 
and above $p_c$. 

\begin{figure}[t]
\psfig{file=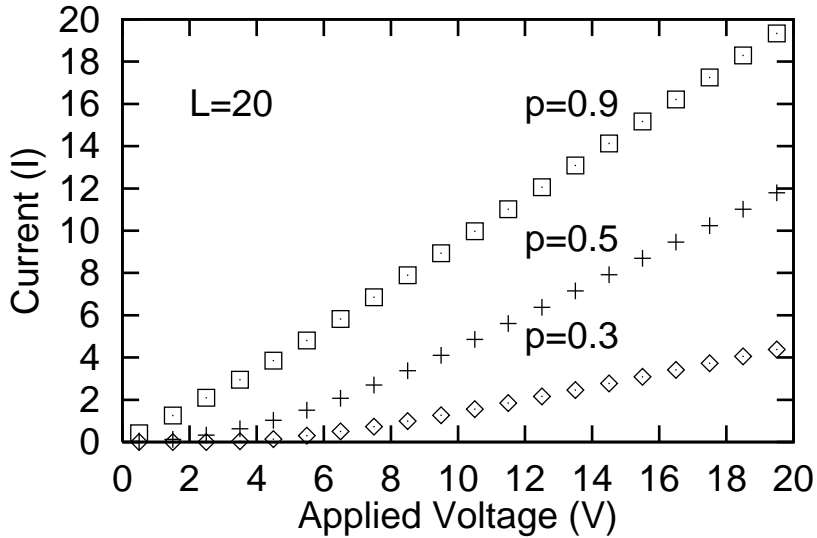}
\caption[]
{Current ($I$) against Voltage ($V$) curves for different volume fractions ($p$)
of the conducting components.}
\end{figure}

The differential conductance $G$ ($\equiv dI/dV$) of the network is 
obtained directly from the $I-V$ curves. A typical such $G-V$ curve is 
shown in {\bf fig.~2} for $L=20$ and $p=0.8$.
To understand the conductance behaviour for the entire network we adopt a 
pedagogical approach where we analyse the elementary prototype circuits with
nonlinear resistors.
The conductance ($G$) of these elementary units 
grows nonlinearly with the applied voltage $V$ and gives us an idea of what
type of functions may be used to fit the $G-V$ data for the much more complex 
macroscopic system. After sifting through various such functional froms, we
find that the simulation 
data obtained through our model system in 2D were best fitted with:
\be
G = G_0 + G_d [1 - exp(-\lambda\Delta V^{\mu})]^{\gamma},    \label{gfit}
\ee
where $G_d = G_f - G_0$ and $\Delta V = V - V_g$, where $V_g$ is the macroscopic
threshold voltage and is the same as $V_B$ above. 
$G_0$ is the conductance in the limit $\Delta V \rightarrow 0$.  Experimentally $G_f$
may be obtained by applying a large enough voltage ($V_s$) such that Joule
heating remains unimportant. 
In our computer simulation on finite sized systems, 
we find $V_s$ to be many orders of magnitude larger than $V_0$ and  $G_f$ is
the conductance when all the tunnelling bonds take part in the conduction. 

\begin{figure}[t]
\psfig{file=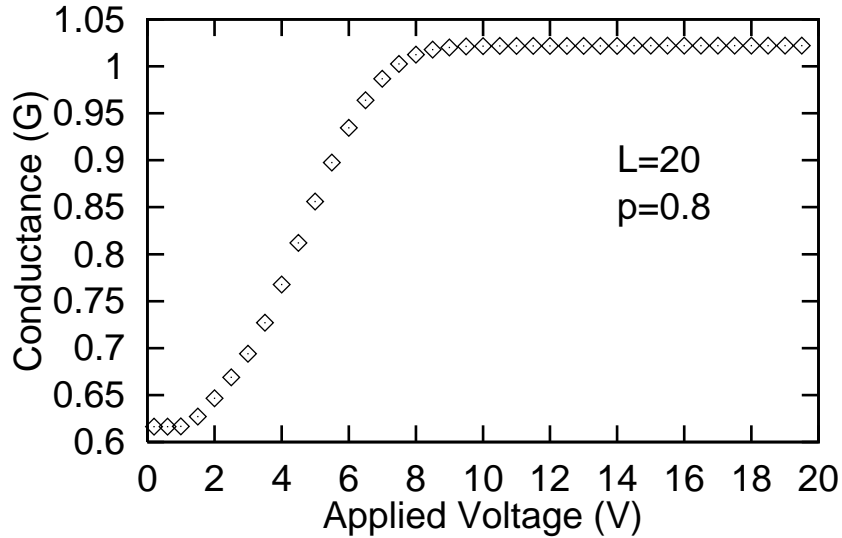}
\caption[]
{A typical curve showing the behaviour of differential conductance $G$ 
against $V$.}
\end{figure}

For a meaningful comparison of all the $G-V$ data with different $G_0$, 
$G_f$, $V_g$, {\em etc.}, we scale the conductance $G$ as $\tilde{G} = 
(G - G_0)/G_d$ and the voltage $V$ as $\tilde{V} = (V - V_g)/V_g$. 
In fact, we tried to scale the $G-V$ data for a set of $p$ in the 
range $0.48 \le p \le 0.52$ ({\em i.e.}, both below and above $p_c$), and
we found that all the data do reasonably collapse. This suggets the following 
general form for the functional behaviour close to $p_c$;
\be
\tilde{G} = f(\tilde{V}),  \label{scaling}
\ee
where $f(x)$ is a function such that $f(0)$ = 0, and $f(\infty)$ = 1.
Here we point out that the threshold (or the breakdown) voltage
$V_g$ is the only relevant voltage scale that enters into the 
scaling function. The other voltage scale $V_s$ 
is seen to have no role in the above scaling
eqn.~(\ref{scaling}).  Expanding eqn.~(\ref{gfit}) near the
onset of nonlinearity ($\Delta V \rightarrow 0$), the excess conductance
$\Delta G = G - G_0$ varies with
the voltage difference ($\Delta V$) as a power-law:
\be
  \Delta G \sim \Delta V^{\mu\gamma} = \Delta V^{\delta},    \label{delta}         
\ee
and the nonlinearity exponent $\delta=\mu\gamma$.  
For $p$ close to $p_c$ ($0.48 \le p \le 0.52$) we find that $\delta \cong 1.0$.
Thus the nonlinearity exponent for the $I-V$ curve is 
$\alpha = \delta + 1 \cong 2.0$.
Experiments in 2D arrays of normal metal islands connected by small tunnel
junctions by  Rimberg {\em et al.} \cite{exp} found $\alpha = 1.80 \pm 0.16$;
suggesting a good support for our model.

Our analysis of results for widely different volume
fractions indicate that the nonlinearity exponent $\delta$ increases
significantly as we go sufficiently away (both below and above) from the
percolation threshold. The scaled data for all the
curves now do not fall on top of each other indicating that all of them can 
not be described by the same fitting function $f(x)$ or by the same fitting
parameters $\mu$ and $\gamma$. Hence
the possible power-law in the regime ($\Delta V \rightarrow 0$) for the onset
of nonlinearity for these curves of different $p$ are {\it not} all the same.

Now the difference between two limiting conductances,  
$G_d = G_f - G_0$ may be
taken as a measure of {\it overall nonlinearity}, whereas the nonlinearity
exponent ($\delta$ or $\alpha$) gives a measure of the {\it initial 
nonlinearity} near the threshold. $G_d$ as a function of $p$ shows a peak 
at around $p = p_c$. So we find that the 
overall nonlinearity is maximum near the geometrical percolation threshold.
Next we looked at how $G_d$ is related to the initial conductance 
$G_0$ in the interval $p_{ct} < p < 1$ in the limit
$L \rightarrow \infty$. The relationship is linear which actually means that 
$G_f$ is also linearly dependent on $G_0$. This in turn implies an identical 
$p$-dependence for the two saturation conductances $G_0$ and $G_f$ around 
their respective thresholds ($p_c$ and $p_{ct}$), consistent with 
the fact that the system has the same conductivity exponent 
in both the zero and the infinite voltage limits ({\em i.e.},
$G_{0,f} \sim (p-p_{c,ct})^t$).

The {\bf AC-response} of the model system also turns out to be very 
interesting. In this case the tunnelling bonds in RRTN are assumed to behave 
as capacitors. The AC-conductance is now expected to behave nonlinearly between 
two saturation regions of $\omega \rightarrow 0^+$ and $\omega \rightarrow 
\infty$ as in the case of DC-response discussed above.
We first give here the EMA where each tunnelling bond has the conductance
$g_{tun} = i\omega c$, where $i = \sqrt{-1}$,  $c$ is the capacitance of the
tunnelling bonds and 
$\omega$ is the circular frequency of the applied sinusoidal voltage with unit 
amplitude. Here we take $c$ = 1 for convenience, thereby setting the frequency 
scale. So for a square lattice ($d = 2$) if we take $g_{ohm} = 1$ in 
eqn.~(\ref{emaeq}), the real part of $G_e(\omega)$ can be shown to be
\be
Re G_e(\omega) = {(2P_{ohm} - 1) \over 2} + {1 \over 2}(X^2 + Y^2)^{1/4}
\cos{\theta \over 2},     \label{recon}
\ee
where $X = (2P_{ohm} - 1)^2 - \omega^2(2P_{tun} - 1)^2$ and 
$Y = 2\omega[(2P_{ohm} - 1)(2P_{tun} - 1) - 2(2P_{ins} - 1)]$ and 
$\theta = \tan^{-1}(Y/X)$. 
It may be checked from the above eqn.~(\ref{recon}) that at $p = p_c$
(= 1/2 in 2D) and in the limit $\omega \rightarrow 0$ the real part of the
complex effective conductance behaves as $ReG_e(\omega) \sim 
\omega^{0.5}$. This is also true for 3D. 

Next we look at the simulation results for the AC-response. It has been
observed that for frequencies $\omega < \omega_0$, one gets some
generic linear or quadratic dependences on $\omega$ which may be
easily understood.  But, for frequencies
$\omega > \omega_0$, we expect percolative effects to gain control and 
$G_{rms}(\omega)$ to follow an equation similar in form to that used for 
the DC-conductance:
\be
G_{rms}(\omega) = G_{rms}(\omega_0) + G_d(\omega)[1 -
\exp(-\lambda[\omega-\omega_0]^{\mu})]^{\gamma}.
\label{acfit}
\ee
For many practical situations, the intermediate frequency range (between 
$\omega_0$ and the upper saturation) is of
prime interest.  In this case, fitting the average graphs, 
we find that $\delta^{\prime} (= \mu\gamma)$ has a minimum value of
about 0.7 near $p_c$, and increases on both sides of it. 
In other words, the AC nonlinearity exponent $\delta^{\prime}$ 
(away from $p_c$) is also $p$-dependent. Notwithstanding this fact,
experiments \cite{carbon, hop} on a wide variety of disordered systems
observe $\delta^{\prime} \cong 0.7$ which matches
closely with our result. 

The behaviour of phase-angle (sometimes called the loss-angle) of the complex 
conductance $G_e(\omega)$ with respect to frequency ($\omega$) is of practical
interest. The phase-angle ($\phi$) is defined through $\tan{\phi} = 
ImG_e(\omega)/ReG_e(\omega)$. The shift of the peak value of it with the 
dilution ($p$) is worth noting. The agreement between the simulation result
and that by EMA is reasonably good. 
The variation of the phase-angle ($\phi$) with frequency ($\omega$)
has been observed for a range of values 0.3 $< p < $ 0.7.
The peak value $\phi_m \cong 0.7$ (radian) is obtained for $p$ around $p_c$
which is close to the universal phase-angle value of $\pi/4$ obtained in the
simple RC model in 2D at $p_c$ predicted by Clerc {\em et. al.}
\cite{clerc}.
We looked at the phase-angle versus frequency for
$p \cong p_c$, and we find therefrom that $\phi_m \cong
0.7$ in this model too.  

Composite systems have very interesting {\bf temperature dependent conduction} 
properties \cite{temp} particularly in the low-temperature regime. 
Some recent experiments
on them show the analysis of their low-temperature data which seem to be 
confusing and contradicting each other. 
The controversy, as briefly described below, is still on and the complete 
physics is yet to be understood. 
The usual attempt is to fit the low-temperature data for such systems by 
the well-known Mott variable range hopping (VRH) formula or with any of its
many generalized forms.
In a very recent experiment by Reghu {\em et al.}, \cite{temp} 
in proton-doped polyaniline networks, it was found that the exponent in VRH
systematically increases from 0.25 to 1 upon decreasing 
the volume fraction $p$ of the conducting component.
Here our goal is not to explain the recent experimental results exactly. 
Rather, our modest hope is to demonstrate
the fact that if one represents the low-temperature data in such 
systems by the VRH or any of its generalized forms, 
then the the relevant exponent in that can change continuously with 
dilution. The approach is again based on percolation theory where we assume the
the activated behaviour for the tunnelling bonds and the metallic behaviour 
to the ohmic bonds. The effect of dilution of the temperature dependent 
conductance behaviour can thus be understood at a preliminary level 
\cite{kgs-temp}.

\vskip 0.3in
\noindent{\large{\bf Discussion}}
\vskip 0.2in

In this report we have discussed various aspects of the nonlinear response in 
the disordered binary composite systems in general.
We have proposed a very simple and minimal model in order to understand 
the nonlinear electrical response and associated physics in composite systems.
In many other physical systems, the response is
negligibly low (or there is no response at all) until and unless the 
driving force exceeds a certain threshold value.
So a class of problems exist where sharp thresholds to transport occur. The
examples in the electrical case is a Zener diode and in the fluid
permeability problems a Bingham fluid (where there is a critical
shear stress $\tau_c$, above which it has a finite viscosity and below
which it is so enormously viscous that it does not flow).
So all these problems may be treated in a similar footing with the 
underlying percolation geometry.

\end{document}